# Comment: Microarrays, Empirical Bayes and the Two-Groups Model

**Yoav Benjamini**

Efron has given us a comprehensive and thoughtful review of his approach to large-scale testing stemming from the challenges of analyzing microarray data. Addressing the microarray challenge right from the emergence of the technology, and adapting the point of view on multiple testing that emphasizes the false discovery rate, Efron's contributions in both fields have been immense. In the discussed paper he reviews philosophy, motivation, methodologies and even practicalities, and along this process gives us a view of the field of statistics from an eagle's eye.

A thorough discussion of such a work is a major undertaking. Instead, I shall first comment on five issues and then discuss new directions for research on large-scale multiple inference that bear on Efron's review. The scope of the review challenges a discussant to try and address some of the issues raised from a broader point of view. I shall give it a try.

## 1. FDR AND LOCAL FDR

Efron notes some of the practical difficulties with the local version of FDR that relies on densities: densities are more difficult to estimate, with higher variability and stronger reliance on assumptions about the tails. These difficulties are even more pronounced in the far tails, where the estimation of the null is more problematic, yet this is where they are usually calculated.

Not surprisingly, I prefer to report and control the FDR $= E(V/R)$, using the usual notation of $R$ being the number rejected, $V$ the number falsely rejected and treating the ratio as 0 when $R = 0$. But my reason is not technical, in the sense that working with cumulative distribution function is easier than with densities, but a fundamental one: the concern to assure reproducible results in the face of selection effects. In other words, how is our inference affected by the fact that we only select as discoveries genes passing some (data-dependent) threshold? Will the identification of discoveries be reproducible when another experimenter will look at a similar question using new data?

*Y. Benjamini is Professor, Department of Statistics and Operations Research, Tel Aviv University, Tel Aviv, Israel e-mail: ybenja@post.tau.ac.il.*

The local false discovery rate fdr($z$) does not cater to the selection process, but rather to the observed value of the statistics and to the abundance of potential discoveries in the pool studied. If it is then used for selection, say by identifying all genes with local FDR values above a threshold, or by displaying all such voxels on a map, the property of the set of genes selected this way is unknown.

Such tail-selection is not only evident in the way results are reported in the examples discussed by Efron, but is needed in order to address the replicability of the results. Identifying a specific gene with a small fdr($z$), it is unlikely that the result will be replicated in another study, in the sense that the fdr($z$) for that gene will be similar. One can still require that the result in the repeated experiment be at least as extreme as the one in the original experiment, possibly relative to a common standard such as the 0.05.

Both arguments call for the evaluation of tail probability (or expectation), thereby considering the effects of selecting all results that are more extreme than a (data-dependent) threshold. Therefore, even in the empirical Bayes framework, the assessment of the selection effect should be by the Fdr (and pFDR) rather than with the local FDR.

In summary, my position is that for the initial screening of potential genes, for the purpose of creating a pool of candidate hypotheses for further study, and *for reporting the results of an experiment in the literature*, I would strongly argue for the use of tail-based measures such as the FDR. The local FDR may still be useful for the decision-making scientist:

the pool of candidate genes is not the end of the story in an investigation, but rather its beginning. More comparisons, literature survey and database searches are commonly done, before, say, a wet lab experiment is conducted on a few genes. When it comes to making personal decisions as to what leads to follow with more extensive research, within the previously identified set, local FDR about a specific gene can give valuable information together with effects size assessment (see below) and the other relevant data gathered.

## 2. THE FDR, THE pFDR OR THE Fdr

The difference between the Bayesian and empirical Bayesian approach on the one hand, and the frequentist approach on the other hand, seems to surface in the form of emphasis on pFDR and Fdr (and fdr) rather than on FDR. Yet in the often used mixture model for microarray analysis, with $n$ large and $p_0 < 1$, for a thresholding procedure at a fixed threshold,

$$\begin{aligned}
\text{FDR} &= E(V/R) \\
&= E(V/R | R > 0)\Pr(R > 0) \\
&\approx E(V/R | R > 0) \\
&= p\text{FDR} \\
&\approx E(V)/E(R) = \text{Fdr}.
\end{aligned}$$

The difference emerges as we deal with problems where $p_0 = 1$ is a real possibility: a trait is not related to any genetic factor, say. In this case the control of FDR is similar to the control of $\Pr(R > 0)$, which in this case is the probability of making any error, while the other concepts are identically 1. So the original FDR, interpretable as a Bayesian concept in one situation, turns out to be the classical frequentist familywise error-rate in another.

The frequentist approach emerges to be useful in yet another important situation, that of estimating sparse signals. As Abramovich et al. (2006) show, in the case when the number of tests grows to infinity, it is optimal in a minimax sense to first use FDR controlling testing and then estimate the significant parameters, thresholding the others to 0. In fact, even when $p_0 = 0$, that is, all parameters are different from 0, but the size of the ordered parameters decays quickly to 0, such FDR-testimation is optimal.

## 3. ESTIMATING THE PROPORTION OF NULL HYPOTHESES

The first versions of our work on FDR included an estimator of the proportion of null hypotheses $p_0$. The inability to publish this work, which had but a single simple theorem and many simulations, led us to drop the adaptive stage where $p_0$ was estimated and replace it by 1, thereby enabling us to get the elegant proof. I believe the original editors did us a favor by requiring mathematical serenity, which led to Benjamini and Hochberg (1995). But they erred when they consequently still refused to publish both our original adaptive results where $p_0$ was estimated, as well as those of Williams, Jones and Tukey, for too strong a reliance on computer work (both appeared in 2000). I think that the right mixture of the two is needed in statistics, as exemplified in Efron's own work.

In the density mixture model it is usually beneficial to estimate $p_0$. Elsewhere, it may be more useful to bound the number of extremely small parameters, rather than to estimate the number exactly at 0: it does not make sense to estimate $p_0$ in the sparsity model from Abramovich et al. (2006), where no parameter is at 0; still it may turn out beneficial to use a bound on the number of parameters nondistinguishable from 0. In the two-stage procedure of Benjamini, Krieger and Yekutieli (2006) such a bound is offered by $n - R(1-q)$. In the adaptive step-down procedure described there, once rejecting $i - 1$ hypotheses, a new bound on $n_0$ is offered by $n_{0i} = n + 1 - i(1-q)$. The procedure steps from $i = 1$ on, as long as $p_{(i)} \le qi/n_{0i}$. The FDR controlling property of this procedure is given in Gavrilov, Benjamini and Sarkar (2008), and its asymptotic optimality follows from Finner, Dickhaus and Roters (2008).

## 4. EFFECT OF DEPENDENCY

There is a misconception that pFDR and local Fdr do not require independence while FDR does. Quoting Efron, false discovery rate control is verified for the procedure in BH "under the assumption of independence among the $N$ $z$-values (relaxed a little)," and that this seems fatal for microarray applications. Yet on the other hand, "a great virtue of the empirical Bayes/two-groups approach is that independence is not necessary."

First, as noted before, in the asymptotic mixture model advocated for microarray analysis, FDR =

$p\mathrm{FDR} = \mathrm{Fdr} = E(V)/E(R)$ as $n$ tends to $\infty$. Obviously the last measure is not sensitive to dependency when a fixed threshold is used, meaning that in this model so will the FDR. Moreover, as noted by Efron, both approaches rely on the same statistic, namely on $\widehat{\mathrm{Fdr}}(z)$, so how can dependency be fatal for one and unnecessary for the other?

Away from the mixture model, the measure $\mathrm{Fdr} = E(V)/E(R)$ is indeed not sensitive to dependence when evaluated at a fixed threshold, unlike FDR or pFDR that summarize the distribution of $V/R$ displayed in Figure 6. But that comes at the expense of destroying the dependence between the number of false discoveries $V$ and the number of discoveries $R$ within the same experiment, as expectations are taken separately. Furthermore, even if the Fdr itself is not sensitive to dependence, its estimator may very well be, and the properties of procedures that take the maximal number of rejections subject to an estimated false discovery measure less than some threshold are prone to have the difficulties mentioned about the procedure in BH under unusual dependency (as in Benjamini and Yekutieli, 2001).

It was mentioned that the procedure in BH controls the FDR under positive regression dependency structure. Even outside this realm, numerous studies indicate the FDR controlling procedure in BH has a robust behavior under the dependency encountered in practical problems. In a systematic study, using a combination of simulations and analytic results, Reiner-Benaim (2007) showed that for two-sided normally distributed test statistics the FDR is always controlled at the desired level $q$ under a wide collection of correlation structures. In extreme situations the FDR may get somewhat higher than $qp_0$, though, so adaptive methods with estimated $p_0$ are somewhat more sensitive to dependency. Interestingly, the structure of constant correlation of, say, all comparisons with same control, where the FDR of the procedure in BH is assured to be less than $q$, is not covered by the current asymptotic results for the pFDR or local FDR. These results require consistency of the empirical distribution of the $p$-values as the number of hypotheses tends to infinity.

In summary, the appropriate statement regarding dependence should be much more balanced than the simple statement that FDR has a problem under dependency and Fdr and local FDR do not. Both approaches are quite robust to the dependence structures encountered in microarray studies, and both are more sensitive when estimators of $p_0$ are incorporated, but not in a critical way.

## 5. ESTIMATING THE DISTRIBUTION UNDER THE NULL

A central theme in Efron's work is the opportunity that large problems offer for estimating the components of the statistical model that are usually treated as assumed. In particular Efron emphasizes rightfully that in many microarray datasets the distribution of the $p$-values evaluated under the assumed null distribution is not uniform, as can be seen either directly or from the nonnormality of the $z$-transformed $p$-values. They are in fact far from normal even in the center, where they should mostly come from the true null hypotheses. Four possible reasons for the discrepancy are discussed. Motivated by the empirical Bayes approach, the estimation of the distribution under the null is offered as a remedy. This remedy may prove useful for frequentist analysis as well.

I agree to all four sources of problems offered by Efron, but would like to offer a fifth one: the set of $p$-values reaching the stage of statistical analysis has been selected from the set originally measured. I noticed this phenomenon a while ago, and commented in a highly prestigious genomics conference that this is not an innocent act. I was almost booed, and the impact of the rest of my lecture diminished, I am afraid. But then, take the microarray examples discussed by Efron: for the three microarray datasets I tried to trace back the reason for the particular number of genes reported in the analysis.

For the HIV data example, the Methods section in the original publication explains: "We used a standard deviation threshold of 50 expression units to select the most variable transcript sequences." For the Singh et al. (2002) prostate data, "Genes whose expression varied less than 5-fold between any two samples in any given experiment were removed." In the BRCA dataset the selection is even more severe. There were 5361 unique genes measured; only 3226 are analyzed. The reason: "In the analyses involving cDNA microarrays, a total of 3226 genes with an average intensity (level of expression) of more than 2500 pixels among all samples, an average spot area of more than 40 pixels, and no more than one sample in which the size of the spot area was 0 pixels were included."

In the first two cases the effect of the selection is clearly to omit more genes with no differences between the groups. Even in the third case, the selection was not (the possibly legitimate) conditioning on the average of normal variates before testing their differences, but a more complicated selection on the original and nonnormal scale.

I suspect that the above reason is as important as the other four proposed for the discrepancy between the real distribution and the assumed one. Worse than that, it affects the center of the distribution under the null usually in a way that distorts the connection between the center and the tails. Therefore inferring about the distribution under the null at the tails from the central part of the empirical distribution is very precarious. For examples of such effects, notice the dips at the center of the actual distributions relative to the estimated distributions under the null in Figure 4(a), and relative to the theoretical null in Figure 1(a).

I do not have a full answer to this difficulty. With some datasets we found that careful preprocessing solved much of the problem. In an extremely large and complicated problem we still struggle, starting from all measured expression data. My point is that practically, certainly with microarray data, I am still more comfortable using an appropriately verified theoretical null distribution than an estimated one. I shall be more confident about estimation if it is tailored to handle the effects of the selection process that hides behind the regular technical microarray preprocessing analysis, a step usually masked from the statisticians' eyes.

With this I end my comments about some of the methodologies offered and opinions expressed, and turn to comment on the future of multiple hypotheses testing in large genomic problems.

## 6. CONFIDENCE INTERVALS FOLLOWING SELECTION

This issue is rarely addressed in the large significance screening studies, so I am happy to find it emphasized in Efron's paper. Too often the decisions as to what clues to follow with expensive research are based on significance only, rather than on estimated effect sizes. The latter calls for making confidence statements about the few selected parameters.

I do not necessarily find here a possible clash between the frequentist and the empirical Bayes approaches. The optimality result in Benjamini and Yekutieli ([2005](#)), which is being viewed as evidence that the two approaches are on clashing orbits, is stated only for two-sided symmetric and equivariant confidence intervals (having the same shape under translation and reflection). Frequentist confidence intervals need not be equivariant. Allowing such flexibility, a confidence statement with a special role for 0 is not a result of Bayesian analysis only: a confidence interval that includes both 0 and an interval not connected to 0 may emerge as a result of inverting nonequivariant acceptance regions, as shown and discussed in Benjamini, Hochberg and Stark ([1998](#)).

Interestingly, in his recent talk in MCP2007, Yekutieli presented a case where Bayesian intervals constructed to incorporate the selection effect used, enjoy False Coverage Rate properties. This indicates that there is potential benefit for successful research on setting confidence intervals after selection, and pursuing this goal from all approaches, frequentist, Bayes, and empirical Bayes, hold better promise for rapid developments, as was the case for testing.

## 7. THE TRANSITION FROM VERY LARGE TO HUGE PROBLEMS

I cannot agree more with Efron when he states that applications is one of the three fundamental forces influencing statistics. Our motivation in developing FDR has been the analysis of clinical trials in which there were 100 or so endpoints. The FDR criterion turned out to be inherently scalable, in the sense that it has stood up to the challenges of tens of thousands of hypotheses. It is becoming more common now to search over millions of hypotheses looking for discoveries against the noisy background (see below), and the FDR is still relevant—possibly because of its triple Frequentist/ Empirical Bayes/Bayes interpretation.

Still, the tools developed along with the approach may have reached the stage where it is unlikely that further polishing of same tools will be of much help: for example, better estimation of $p_0$ will offer little improvement, because in such huge problems $p_0$ is very close to 1 or even 1; the discoveries are not likely to be important if abundant in the extremely large pool searched. That the tools are polished is in fact a tribute to the many researchers in the statistical community who in their (sometimes competitive) efforts advanced tremendously our knowledge and understanding about false discovery rates.

There are three possible directions to deal with the new challenges of huge multiple testing problems using new tools, and almost all are related in some sense to the approach presented by Efron.

D1. The enrichment analysis offered by Efron points at an important first direction: increasing signal to noise ratio by collecting hypotheses to sets in which they are likely to be true together, or false together. In Efron's gene-enrichment example the clustering of genes into sets is based on external information regarding the pathways involved. In the brain-imaging example the clusters are based on a moving window (for FDR controlling scan statistics see also Pacifico et al., 2007). The cluster-based analysis in Benjamini and Heller (2007) is another example of the enrichment approach in the brain-imaging problem. It comes in a different flavor, though: based on the pilot study routinely performed in brain-imaging experiments, the voxels in the brain are first clustered to create a coarser partition of the hypotheses. The clusters need not be of the same size and shape, and are of neurological relevance. Then, in the main experiment, the clusters are tested using a combining statistic for each cluster. Therefore, not only do we gain increased power from combining the evidence over clusters, but also when addressing multiplicity the number of clusters tested is much smaller than the number of voxels tested.

The essence of the above examples is clear. When the tested parameters have further structure, in the sense that we can have a grasp in what sets the hypotheses are going to be true together and false together (correlated parameters), enrichment analysis is of great potential: in many cases not only is the signal to noise ratio increased but the multiplicity problem can be reduced.

D2. The second direction is that of employing weights to differentiate between the hypotheses tested. The weights may incorporate differing importance of the hypotheses (Benjamini and Hochberg, 1997), or different prospects for showing effects (Genovese, Roeder and Wasserman, 2006). As in the case of enrichment analysis, the weights can be based on outside information, or on information from initial testing. Weights can also answer Efron's third concern regarding the enrichment analysis, especially as one may assign weights to sets of hypotheses in a way that is proportional to the size of the set, as is done in Benjamini and Heller (2007).

D3. The third direction is that of endowing a hierarchical structure to the family of hypotheses tested, where a subfamily of hypotheses at a branch is tested only after the node from which it branches is tested and rejected. When such hierarchical structure enjoys an enrichment property, again in the sense that the hypotheses in a branch tend to be true or false together, an opportunity for power gain arises. Reiner et al. (2007), for example, test the association between the expression of some 27K genes in each of five brain regions, and 17 behavioral traits, a study of more than 2.2 million hypotheses. They first screen for genes and brain regions where strain differences exist. Only those combinations of brain regions and genes passing an FDR-based threshold are further tested for correlation, each subfamily of 17 tests of correlation being tested separately. The theoretical questions regarding such procedures are discussed in Yekutieli (2008), and answers are given there for the case where the test statistics at a node and at its branching hypotheses are independent. In Benjamini and Heller (2007) described above, a natural hierarchy is to test clusters of voxels and then individual voxels within clusters. The test statistics for testing a cluster and those for testing voxels within a rejected cluster are not independent, so conditional $p$-values have to be estimated for their hierarchical use.

Recent work by Meinhausen (2008) for testing the importance of a large number of variables in a regression model also makes use of a hierarchical approach, this time within the familywise error-rate framework. As to the design questions for very large experiments raised by Efron, they can also be answered within a hierarchical setting. Zehetmayer, Bauer and Posch (2005), for example, first use a screening experiment, where each hypothesis is tested with no attention to multiplicity, and based on new data conditional $p$-values for the hypotheses that pass the initial screening are calculated and multiplicity is addressed by FDR control.

Most of these efforts seem to indicate that managing to create hierarchical testing structures that enjoy enrichment properties is extremely promising. Tools such as the estimated distribution under the null, the estimated $p_0$ at a branch, and sometimes the local FDR, which emerged from the empirical Bayes perspective, can remain useful but may need further adjustments. Allow me to remain philosophical, at this point, and not dwell on details (especially since I do not know how to handle them).

## 8. IN CONCLUSION

I enjoyed reading the first sections offering a broad view of past achievements, even when I disagree with some of the solutions offered. I am enthusiastic about the last three sections, in which new directions of progress are identified to answer needs in applications, as we pass from very large problems to huge ones. I found it important to emphasize the more general nature of these directions, and connect them to current research efforts that reflect similar attitudes. I thank again Efron for taking such a broad view of the subject, thereby calling for a more-than-technical discussion on my part.

## ACKNOWLEDGMENTS

I would like to thank Dani Yekutieli for engaging discussions about the above ideas. This work was supported by grants from GIF, NIH and ISF.